\documentclass[fleqn,10pt]{wlscirep}
\usepackage[T1]{fontenc}
\usepackage{sectsty}
\usepackage{graphicx}
\usepackage{amssymb,amsmath}
\usepackage{wasysym}
\title{Negative Magnetoresistance in Amorphous Indium Oxide Wires}
\author[1,$\dagger$,*]{Sreemanta Mitra}
\author[1]{Girish C Tewari}
\author[1]{Diana Mahalu}
\author[1]{Dan Shahar}
\affil[1]{The Weizmann Institute of Science, Department of Condensed Matter Physics, Rehovot, 76100, Israel}
\affil[$\dagger$]{Current address: Nanyang Technological University, School of Physical and Mathematical Sciences, Divison of Physics and Applied Physics, Singapore, 637371}
\affil[*]{sreemanta85@gmail.com}
\begin{abstract}
	We study magneto-transport properties of several amorphous Indium oxide nanowires of different widths. The wires show 
	superconducting transition at zero magnetic field, but, there exist a finite resistance at the lowest temperature. The $R(T)$ broadening was explained by available phase slip models. At low
	field, and far below the superconducting critical temperature, the wires with diameter equal to or less than 100 nm, 
	show negative magnetoresistance (nMR). The magnitude of nMR and the crossover field are found to be dependent on both temperature and the cross-sectional area. We find that this intriguing behavior originates from the interplay between two field dependent contributions. 
\end{abstract}

\begin{document}

\flushbottom
\maketitle
%
%
\thispagestyle{empty}


\section*{Introduction}
Superconducting materials, due to their multiple characteristic length scales, viz.; penetration depth, coherence length and Fermi wavelength, show
richer phenomena than other non-superconducting materials, in nanoscle dimension \cite{prl109n}.     
The onset of superconductivity is
accounted for by nucleation of the superconducting phase
shunting the normal current. This effect as observed in
superconducting systems, is determined by the temperature dependent coherence length ($\xi$), irrespective of their dimensionality\cite{aru1}.
 At the low temperature part of the transition,
dimensionality plays a crucial role. As soon as, if nothing else, one channel of supercurrent nucleates, in {\it 2D} and
{\it 3D} systems, spots of normal phase do not contribute to
zero DC resistance and temperature dependence of resistance has an abrupt bottom part. However, in {\it 1D}, where the cross-sectional area, $\sigma$ $\leq$ $\xi^{2}$, there exists only one parallel channel of supercurrent\cite{aru1}, phase fluctuations (known
as phase slips) lead to a preterm suppression of the superconducting properties. As a consequence, superconducting nanowires are nonlinear elements 
and are dual to Josephson junctions\cite{mooijnjp} where, the roles of phase 
and charge, and simultaneously current and voltage, are interchanged. Disorder is also capable to tune and induce new correlated
electron physics in low-dimensional materials as observed in a recent work on quasi {\it 1D} superconductors\cite{alexnc}.
The study of quasi-one dimensional ({\it 1D}) superconductors deals with the fundamental question, pointed out
by Little \cite{1little}, of whether there exist a superconducting state, or enhanced fluctuations in low-dimension suppress
the phase coherence. 
As the temperature ({\it T}) drops below a critical value, $T_{C}$, the resistance ({\it R}) in {\it 1D} superconductors drops, but it ceased to go to zero like its bulk counterpart, and a finite {\it R} persists \cite{prb5} to the lowest {\it T}. The 
low energy excitations, leading to this residual {\it R}, represent a local disturbance of 
superconductivity for a short period of time. These fluctuations represents activation process which are triggered by either thermal, where the population 
is strongly {\it T} dependent \cite{la,mh}, or through quantum mechanical tunneling, which have a weak {\it T} dependence \cite{giordanoprl,3giodano}.
\par 
Thermal phase slip was accounted well within the theory developed by Langer, Ambegaokar, McCumber, Halperin (LAMH) \cite{la,mh},
and was confirmed by experiments on 0.5{$\mu$m} tin whiskers\cite{prb5}. The interpretation of quantum phase slip was rather complex. 
Study of such dissipative superconductivity has been a difficult task
as the non-equilibrium quasiparticles are massively generated
by phase slip process \cite{chennp}. If not removed effectively, they tend to overheat the
nanowire, driving it into the normal state. Thin MoGe
\cite{sahunp,shahprl} and Al wires \cite{pengprl,singhprb} seem to be switched into the normal
state by a single phase slip. In an inhomogeneous
superconductor, reentrance \cite{cpacsn} (switching to insulating state again as T$\rightarrow$0) may occur if the Josephson energy
rises with respect to the thermal energy (or the Coulomb energy in granular materials). 
Thermal and quantum phase slips were also observed in  narrow junction of Al, fabricated by controlled electromigration\cite{baunc}.  
 Quantum phase slip phenomenon was also observed in ultranarrow 
superconducting nanorings \cite{arusr} and Josephson junction ring\cite{prb93}. In superconductor nanoring \cite{arusr}, the phenomenon is responsible for suppression of persistent current, another basic attribute of superconductivity. It has been shown that QPS significantly affect their magnitude, the period and the shape of the current-phase relation. 
There have also been some ambiguities in {\it T}-dependence of {\it R} for superconducting wires.
Long {\it R} tail was absent in Pb 
wires as narrow as 15 nm \cite{prl71} and the system went to
a $\textquoteleft$true superconducting state' however, narrow Al wires exhibit the long {\it R} tail\cite{prb77}. Alternatively, in case of previously studied
amorphous Indium oxide nanowires \cite{prl95},  
some wires showed vanishing {\it R} at low-{\it T}, and some saturated at a non-zero value. For titanium nanowires of effective diameter of the order of 50 nm, QPS has significantly broaden the $R(T)$ \cite{lehprb85}. Similar type of broadened transition and {\it R} plateau was previously observed in  MoGe nanowires and attributed to quantum phase slips\cite{lauprl}.
Also a recent work on indium oxide nanowires\cite{mitraprb} showed flattening of {\it R} as {\it T} is decreased and attributed it's occurrence to a possible macroscopic quantum mechanical tunneling of vortices between pinning sites. 
While going from a {\it 2D} to a quasi-{\it 1D} system,  $T_{c}$ is found to be 
decreased in case of amorphous MoGe films as the width
is reduced and superconductivity was completely destroyed
in the 1D limit\cite{dynesprl59}. A recent work suggested that decrease in $T_{c}$ follows
exponential variation with the inverse of the wires’ cross-sectional area\cite{kimprl109}.
\par 
In superconductors, application of {\it B} enhance the superconducting fluctuations through
two effects, by aligning electron spins and by increasing the kinetic energy of the electrons via Meissner screening current \cite{tinkhambook} and as a result {\it R} increase with {\it B}.
Nonetheless, in narrow superconducting strips of Pb \cite{dynesprl} and Al \cite{prb77}, at small magnetic fields, negative magnetoresistance (nMR) is observed.
Out of several non-mutually-exclusive theoretical pictures of nMR \cite{nmrmod1,nmrmod2,nmrmod3}, 
no commonly accepted model for this phenomenon has been found yet and has been an open question in {\it 1D} superconductivity \cite{pr464}.\par 
In this paper, we present the results of  low-{\it T} magneto transport measurements on a series of amorphous Indium oxide (a:InO) nanowires. Pronounced nMR in nanowires 
of width less than or equal to 100 nm, irrespective of length, has been observed. The magnitude of nMR is found to have a size and {\it T} dependence. 

\section*{Results and Discussions}
\begin{figure}[h]
  	\includegraphics[width=8.5 cm]{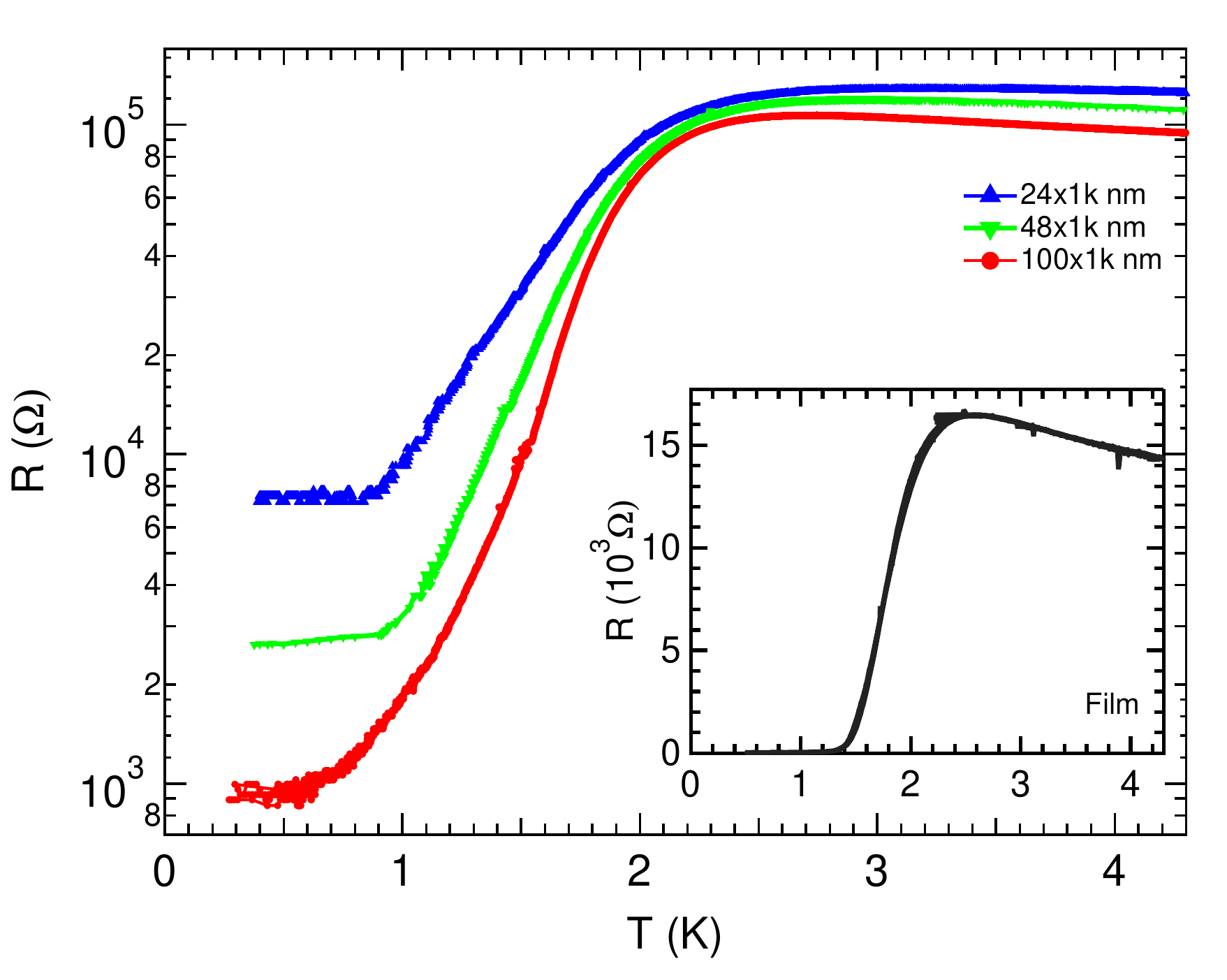}
  	\caption{ The {\it T} variation of {\it R} (in semi-log scale) at {\it B}=0 for 1 ${\mu }$m  
  		long wires with different [24(blue), 48(green) and 100(red) nm] widths. The wires show 
  		superconducting transition at $T_{C}$ around 1.8 K. Below $T_{C}$, at lowest {\it T}, {\it R} saturate at finite values. 
  		(Inset:) {\it R-T} for the simultaneously prepared film, measured at {\it B}=0 shows a sharp drop at $T_{C}$=1.76 K and {\it R}
  		appeared to go to zero.}
  	\label{figrt}
  \end{figure}

We begin to present our findings by showing the variation of {\it R} with {\it T}, measured at {\it B}=0. 
The respective {\it R(T)} plots, for 1 ${\mu}$m long wires with 3 different {\it w}, 24, 48, 100 nm, are
shown in figure \ref{figrt}.  
The co-fabricated reference film has also been measured and the zero field $R(T)$ for it is shown in the inset of figure \ref{figrt} for comparison. 
It is seen that all the 
wires exhibited exponential (linear in semi-log scale) drop, signifying superconducting transition at {\it B}=0. Nonetheless, they saturated at non-zero values and a residual {\it R} persist at low-{\it T} .
 The approach of {\it 1D} regime is herald by 
the higher value of the transition width ($\varDelta T$) in 
wires ($\varDelta T_{wire}\approx$ 2 K) than that of the film ($\varDelta T_{Film}\approx$ 1.10 K) \cite{prb41,prl71}.  
\par 
The presence of residual {\it R} in the wires, far below $T_{C}$, and broadening of $R(T)$ might be related to the 
phase slip process due to quantum fluctuations. 
There can obviously be other source of $R(T)$ broadening that one can 
consider. The widening could result from the fact that a too high current has been used
during the measurements. To rule out this possibility, we acquired $R(T)$ curves for small current value (see Methods\ref{meth}). 
It is also recognized that inhomogeneities may give rise to broad transition\cite{prb75,prb77}. Although our system, a:InO is disordered by default, the film showed a sudden drop at $T_{C}$ signifies that this thickness of evaporated a:InO leads to continuous films and the morphology is not granular\cite{mitraprb}. 
\par
\begin{figure}[h]
  	\includegraphics[width=10 cm]{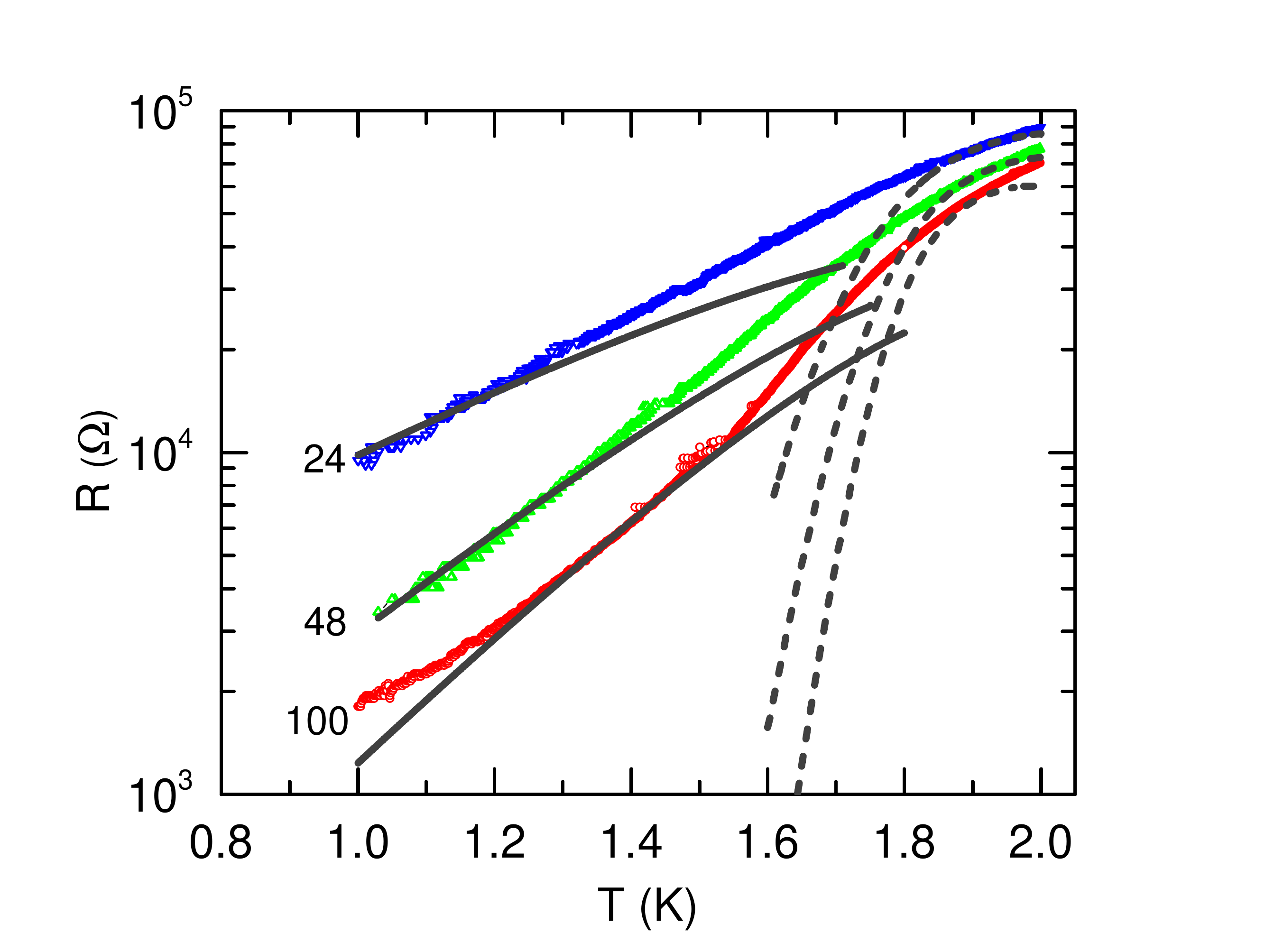}
  	\caption{$R vs T$ for the 3 wires of different widths as in Fig\ref{figrt}. The dashed lines are the fitting of the experimental data with TAPS model, whereas the solid lines are the fitting with the simplified short wire model. }
  	\label{figrtfit}
  \end{figure}

 
A more convincing and plausible way to explain the $R(T)$ broadening and non-zero {\it R} at lowest {\it T}, is the excitation of phase slips. 
When the activation of phase slips is thermally driven, the formula for {\it R} is obtained from a model, originally described by  Langer–Ambegaokar–McCumber–Halperin (LAMH). 
Within this model, phase slip formation necessitates overcoming an energy barrier ($\Delta F$). A characteristic timescale for
the  fluctuations is  fixed by a pre-factor $\Omega$, related to
the attempt frequency of random excursions in the superconducting order parameter.
The resultant  fluctuation-dominated resistance of a {\it 1D} superconducting wire can be expressed as follows
\begin{equation}
R_{TAPS}(T)=\frac{\pi\hslash^{2}\Omega(T)}{2e^{2}k_{B}T}\exp(-\frac{\Delta F(T)}{k_{B}T}) 
\label{eTAPS}
\end{equation}       
where, $\Omega=(\frac{L}{\xi})(\frac{\Delta F}{k_{B}T})^{1/2}(\frac{1}{\tau_{GL}})$ is the attempt frequency, $\tau_{GL}=\frac{\pi\hslash}{8k_{B}(T_{ons}-T)}$ is the Ginzburg-Landau characteristic relaxation time and $k_{B}$ is the Boltzmann constant.  
The experimental data with the LAMH fitting in dashed line is shown in Fig.\ref{figrtfit}. This model can only describe the experimental data near $T_{C}$ and clearly fails as {\it T} is lowered. It is worth noting here that the peak as obtained from eq. \ref{eTAPS}, has no physical meaning and is a mathematical artifact\cite{cpacsn}. 
\par 
The failure of TAPS model at lower {\it T}s, allowed us to consider QPS phenomenon to be present at that side of the data. Because of the high normal state resistance in our sample it is advantageous for us to use QPS\cite{lehprb85}. Following the simplified short wire model, \cite{pu41,zaiprl} the QPS contribution to the
effective resistance of a superconducting wire with length {\it L} and cross section $\sigma$ can be written as, 
\begin{equation}
R_{QPS}(T)=c\frac{\Delta(T)S_{QPS}^{2}L}{\xi(T)}\exp(-S_{QPS})
\end{equation}  
where $\Delta(T)$ and $\xi(T)$ are temperature dependent superconducting energy gap and coherence length, respectively. $S_{QPS}=A\frac{R_{Q}}{\xi}\frac{L}{R_{N}}$ is the QPS action and $R_{Q}=h/4e^{2}$ is the quantum resistance of Cooper pairs, where $h$ is Plank's constant and {\it e} is the electronic charge. The fitting to this form, using {\it A} and $(L/\xi)$ as parameter, has also been shown in figure\ref{figrtfit} as solid lines. The onset temperature and the normal-state resistance can be obtained from experimental {\it RT} dependences. For all wires the best fitted values of {\it A} is in the range of 0.13 to 0.16. We
found that the correspondence between the simplified short-wire model\cite{zaiprl} and the experiment can be considered as good. 
\par
\begin{figure}[h]
   	\includegraphics[width=10 cm]{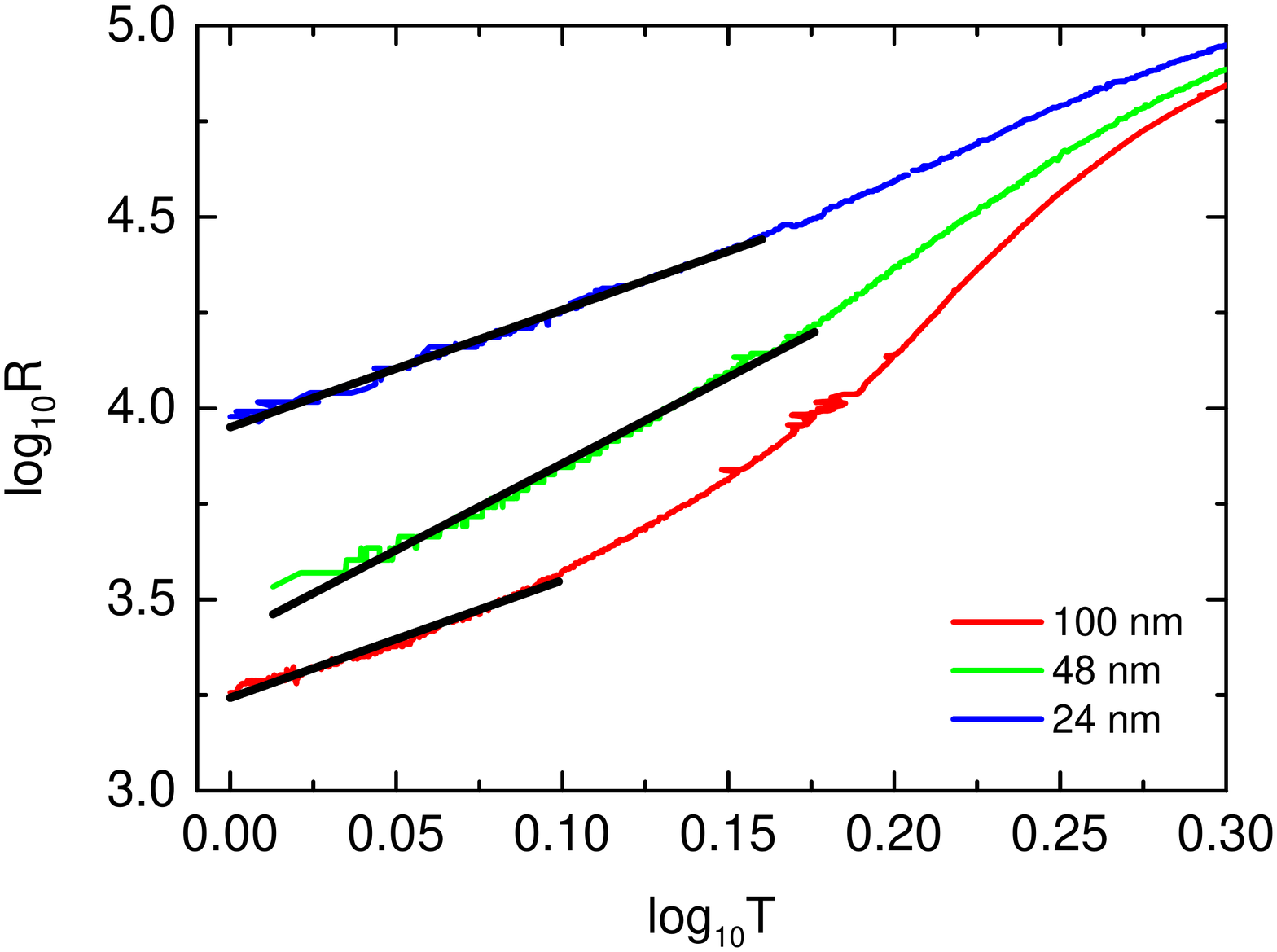}
   	\caption{Resistance vs temperature for three
   		samples of the same length but different width as in Fig. \ref{figrtfit}. Black solid lines are fittings to power dependence $R\sim T^{2\gamma-2}$. }
   	\label{figrtrg}
   \end{figure}

 
We also explore that at certain limits, this renormalization model\cite{zaiprl} predicts a functional dependence of the effective {\it R} on {\it T}.   
In the high-temperature limit, $R\sim T^{2\gamma-2}$, where $\gamma=(\frac{R_{Q}}{R_{qp}})$ is the dimensionless conductance related
to the effective quasiparticle resistance $R_{qp}$, and associated with dissipation provided by the quasiparticle channel\cite{prb77}. 
The results of fitting on our $R(T)$ data with this form has been shown in figure \ref{figrtrg}. The value of $R_{qp}$ came out from the fitting is 2.5 $k\Omega$, which is significantly smaller than $R_{N}$.  
\par 
The saturation in {\it R} at the low-{\it T} is not understood, but it might be due to isolated barriers such as
microcracks and twin boundaries separating macroscopic phase-coherent superconducting regions\cite{cpacsn}. It is interesting to note that $R_{sat}$ increases as decrease in wire width, although at normal state ({\it T}=4 K), all of them had almost same {\it R}. This might indicate the increase of phase slip centers as the width of wires decreases. 
\par
\begin{figure}
  	\includegraphics[width=8.5 cm]{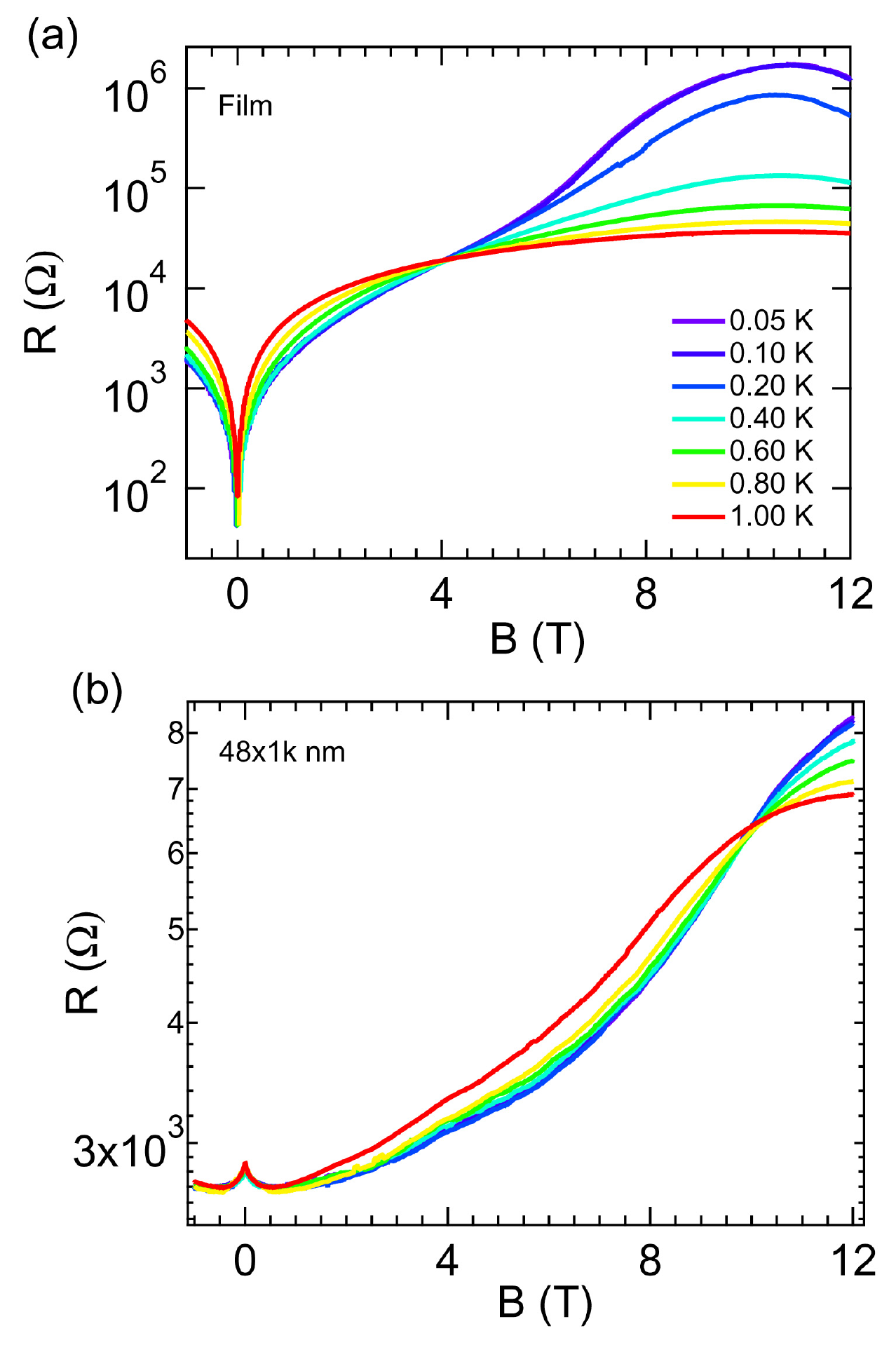}
  	\caption{ (a)The {\it R vs B} isotherms (in semi-log scale) for the film,
  		showing {\it B}-driven SIT ($B_{c}\approx $4.05 T). 
  		At $B_{c}$, $R_{\Square}$ is 6.5 $k\Omega$, which is close to quantum  resistance of Cooper pair. 
  		(b) The {\it R(B)} isotherms for {\it w}= 48 nm wire at the same {\it T}s as of the film (see the color palette).
  		The low-field magnetoresistance peak at {\it B}=0 is clearly visible. The isotherms cross each other at {\it B}=10 T, signifying SIT.
  	}       
  	\label{figrb}
  \end{figure}

For the rest, we will focus on the {\it B} dependence of {\it R} of the devices.
We begin the presentation of our {\it B}-dependent data by plotting, in figure \ref{figrb}(a), the {\it R(B)} isotherms (in semi-log scale) obtained for the film, over our entire {\it B} range. The isotherms crossed each other at a particular {\it B}, 
($B_{c}\cong$4.05 T), signifying SIT. At $B_{c}$, the sheet {\it R} ($R_{\Square}$) is equal to 6.5 $k\Omega$ which is close to $R_{Q} [=6.47 k\Omega$],
and is in accordance with the bosonic description of SIT \cite{fisherprl}. The high-{\it B} phenomenology exhibited by this film was similar to our previously studied  a:InO films \cite{murthyprl1,murthyprl2,maoznat,dannys,mitraprb}.
In fig. \ref{figrb}(b) the {\it R} vs {\it B} isotherms for the 48 nm wide wire is shown.
Although {\it B} is expected to increase {\it R}, we find, near zero {\it B}, the residual {\it R} was actually suppressed by the application of {\it B}, leading to nMR over a certain {\it B} range. After that, {\it R} increased with {\it B}. The {\it B}-driven SIT occurs as the isotherms crossed at 10 T. 
It is interesting to see that, although at zero {\it B} the wire has a higher {\it R} than the film, 
its {\it R} changed much slowly with {\it B} in comparison and did not reach to as high {\it R} as the film at low-{\it T}\cite{mitraprb}. 
The appearance of low-{\it B} nMR in sub-100 nm wide wires is the main focus here.
\par
\begin{figure}
  	\includegraphics[width=8.5 cm]{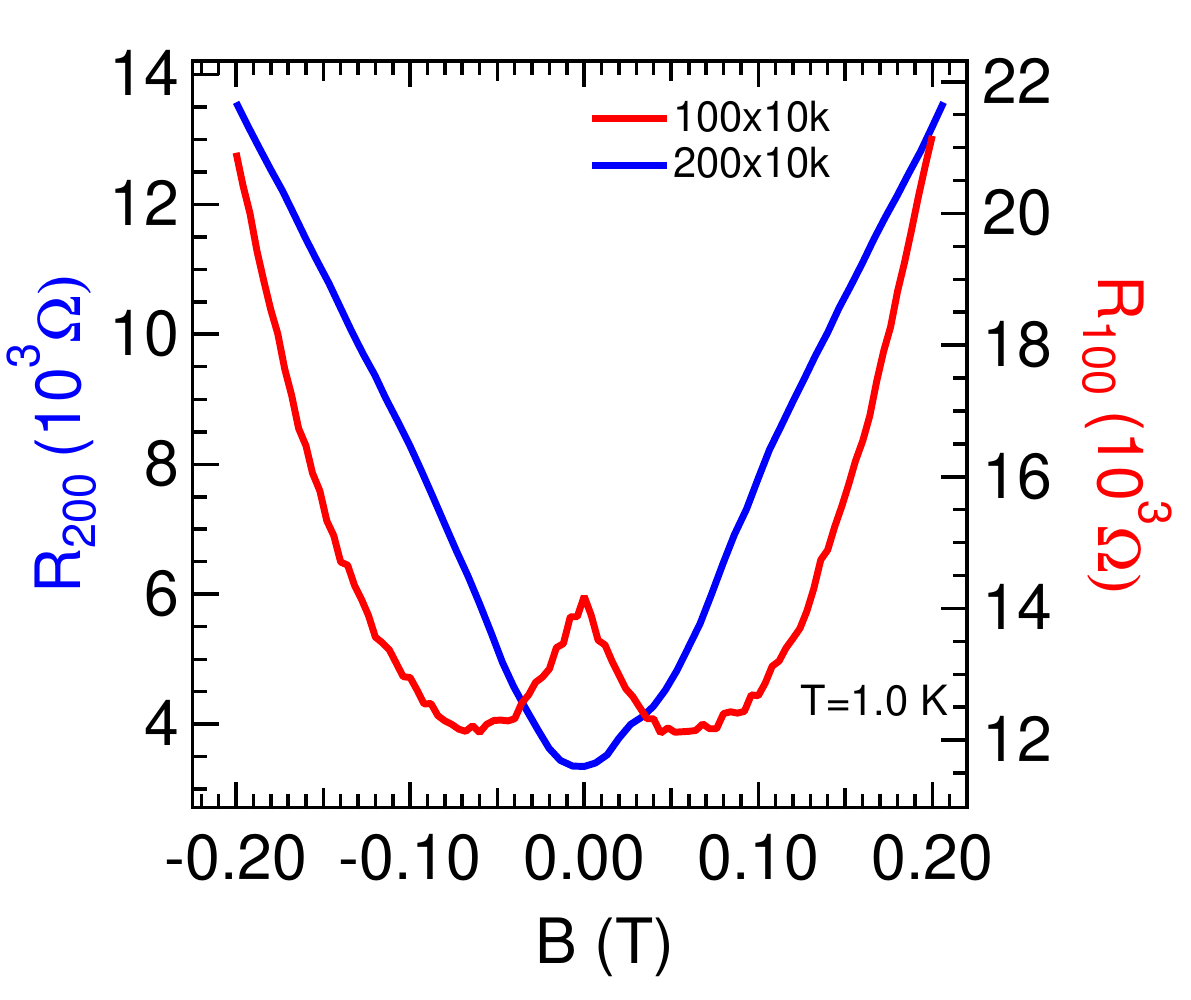}
  	\caption{ Low-field magnetoresistance data for 100 and 200 nm wide wire ({\it L}=10 $\mu$m) measured at {\it T}= 1 K. 
  		The left axis is for 200 nm wire and the right axis is for 100 nm wire. 
  		Negative magnetoresistance near {\it B}=0 is observed in 100 nm wire whereas it is absent in 200 nm wire.}       
  	\label{figrbn}
  \end{figure}

 
In order to establish that the observed nMR appeared in wires of width 
less than or equal to 100 nm, irrespective of length, we show in fig. \ref{figrbn} the {\it R vs B} isotherms for 200 and 100 nm wide
and  10 $\mu$m long wire. The left axis shows the {\it R} value for the 200 nm wide wire, whereas, the right
axis is for the 100 nm wide wire. Negative magnetoresistance near {\it B}=0 is observed in 100 nm wire whereas it is absent in 200 nm wire. 
For 1 $\mu$m long wire with 100 nm width, pronounced nMR was also observed and will be discussed later. 
\par 
We believe that the nMR
is a genuine effect and not an artifact of the contacts, if it was so, the wider wires along with the film would 
also show this behavior. In order to get more insight of this 
effect, we studied its detail {\it T} and {\it w} dependence for {\it L}=1 ${\mu}$m long wires. 
\par
\begin{figure}[!h]
  	\includegraphics[width=8.5 cm]{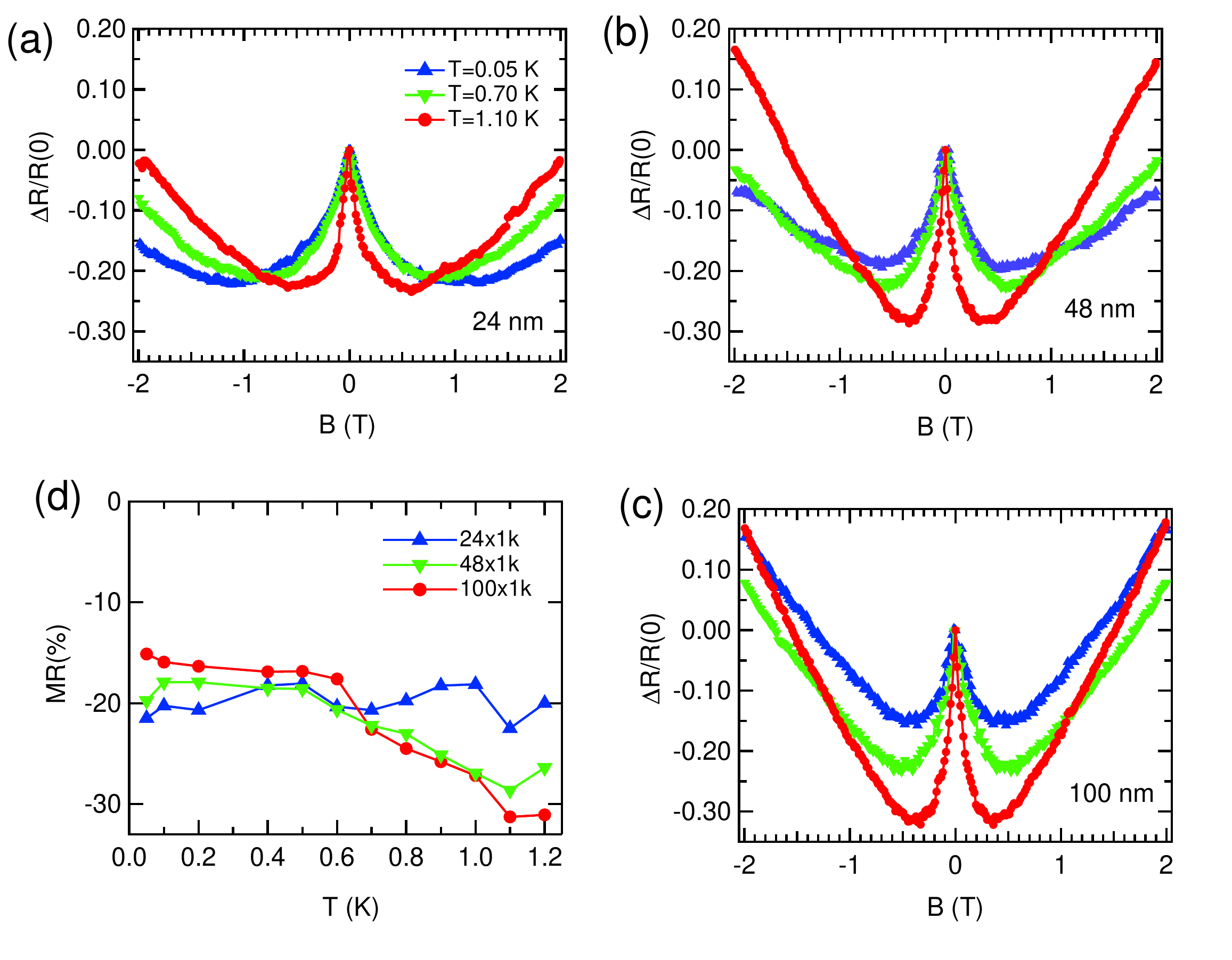}
  	\caption{ The variation of change in {\it R} with {\it B} measured at 3 different {\it T}s;
  		0.05 K (blue), 0.70 K (green) and 1.10 K (red) for {\it w}= (a) 24 nm (b) 48 nm and (c) 100 nm wire. 
  		The data are normalized with respect to zero {\it B} resistance [{\it R}(0)]. $\varDelta R=[R-R(0)]$ is the change
  		in {\it R} from {\it R}(0). The lines joining the data points are to guide the eye.
  		(d) The variation of percentage change of MR [$\frac{\varDelta R}{R(0)}\times 100\%$] with {\it T} for wires of 
  		{\it w}= 24 nm (blue), 48 nm (green) and 100 nm (red). The lines are to guide the eye.}
  	\label{figdRdT}
  \end{figure}

In figures \ref{figdRdT}(a)-(c), the variation of change in {\it R} with {\it B} near zero-{\it B}, measured at 3 different 
{\it T}s, viz; 0.05, 0.70 and 1.10 K  for {\it w}=24, 48 and 100 nm  wires are shown. 
For the ease of comparison, we prefer to present our data in a normalized scale, 
where {\it R}(0) is the zero {\it B} resistance and $\varDelta R=[R-R(0)]$ is the change in {\it R} from
the zero {\it B} value. The wire resistance decreased by 20-30{\%} from their respective zero-{\it B} values,
before {\it R} started to increase. It can be seen from fig. \ref{figdRdT}(d) that magnitude of nMR increased
with {\it T} up to 1.10 K in all the wires. 
The $T_{C}$ of the wires are around  1.8 K, and unfortunately we were unable to measure the 
MR very close to $T_{C}$, where this nMR might disappear. 
The change in magnitude of nMR with {\it T} is more prominent in the wider wires whereas for the narrowest 
of them ({\it w}=24 nm ) it is fairly constant around 20{\%}.
 \par
\begin{figure}[!h]
  	\includegraphics[width=8.5 cm]{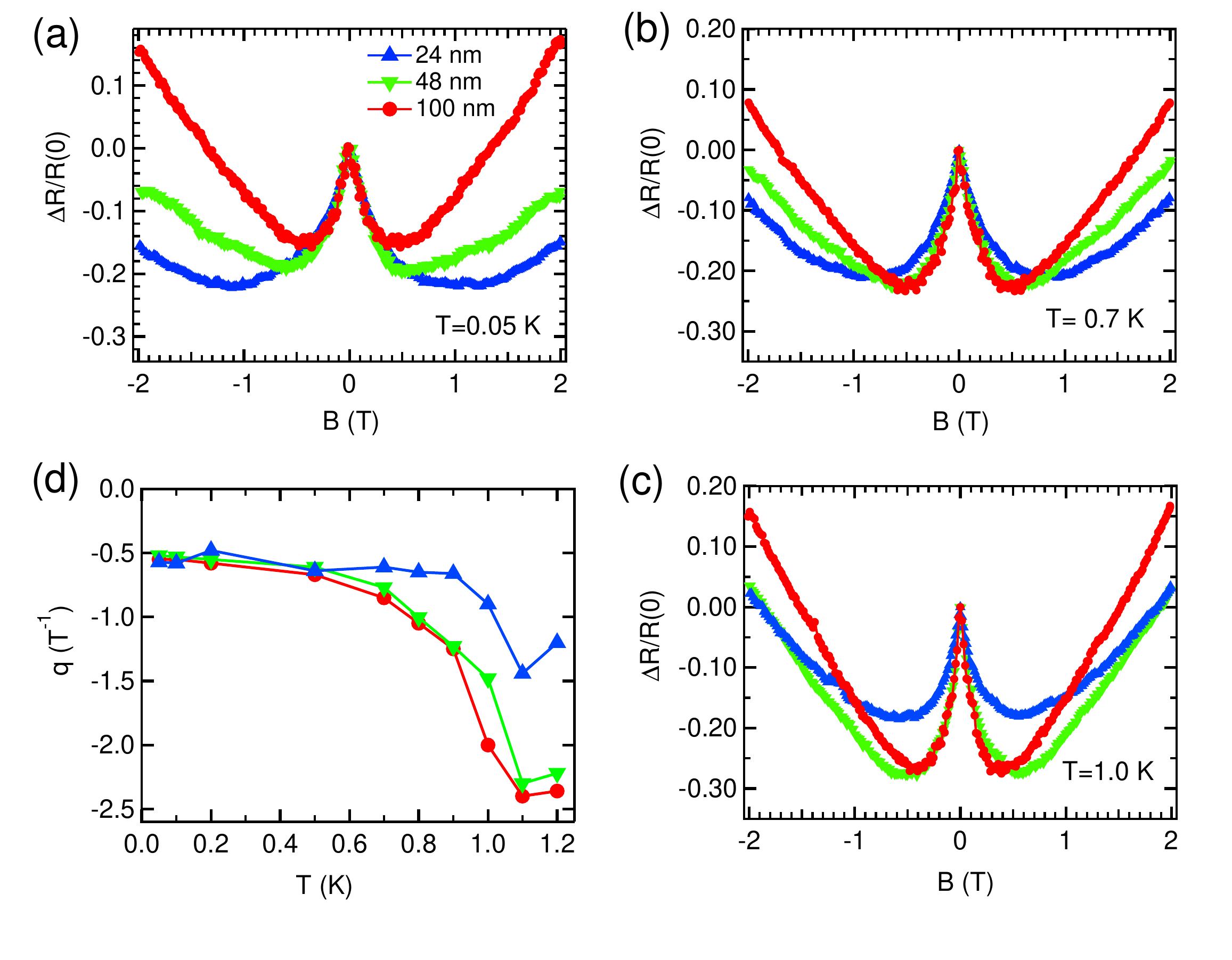}
  	\caption{ The variation of change in {\it R} with {\it B} measured for 3 different {\it w};
  		24 (blue), 48 (green) and 100 nm (red) at {\it T}=(a) 0.05 K  (b) 0.70 K  and (c) 1.0 K. The data are normalized
  		with respect to zero {\it B} resistance [{\it R}(0)]. $\varDelta R=[R-R(0)]$ is the change in {\it R} from {\it R}(0).
  		The lines joining the data points are to guide the eye.(d) The variation of the slope ({\it q}) (see text) with {\it T} 
  		suggests that in the nMR regime the rate of change of {\it R} with {\it B} increases as {\it T} is increased. 
  		For wires of different {\it w}, {\it q} stays fairly constant up to 0.5 K. The lines are to guide the eye. } 
  	\label{figdRdW}
  \end{figure}

 
 To look at the effect of wire width on the nMR, we plot in fig. \ref{figdRdW}(a)-(c) the nMR measured for 3 different widths; 
 24, 48 and 100 nm at 0.05 K, 0.70 K and 1.0 K respectively.
 At low-{\it T} the narrower wire has more change than the wider wires, whereas the scenario changes as {\it T} is increased. 
 It is also seen that at low-{\it T}, all the wires have, almost same broadening and as {\it T} was increased, 
 the wider wires changed more sharply than the narrower one. In order to represent this observation quantitatively, 
 we followed a prescription described hereinafter. Since near {\it B}=0, {\it R} changes linearly with {\it B}, 
 we calculate the slope to measure how fast {\it R} changed with {\it B}. The {\it T} variation of this slope ({\it q}) 
 for wires with different {\it w} is shown in fig. \ref{figdRdW}(d). The 
 slope for all the wires stayed fairly constant
 up to 0.5 K and changed above that. The magnitude of nMR for different wires [figure \ref{figdRdT}(d)] crossed around the same {\it T}.
 The crossover field, the magnetic field after which the positive MR took over, was decreased with increase in {\it T} for all the wires,
 which also came out of this analysis independently. This also suggests that although up to 0.5 K, the crossover field does not depend on size, but it decreased with the increase in wire width above that.  
 \par 
 It is inferred from figures \ref{figdRdT} and \ref{figdRdW} 
 that {\it T} and wire's {\it w} have a close correlation in the evolution of this low-{\it B} nMR 
 in ultra-narrow wires.
 \par 
 One of the plausible explanations of this nMR is related to presence of magnetic impurities
 and subsequent Kondo mechanism. We rule out this possibility based on couple of features. 
 First,  a:InO was evaporated with 
 ultra-pure In$_{2}$O$_{3}$ pellets, in an HV chamber, where magnetic materials have never been processed. Second, 
 there is a pronounced diameter dependence, as discussed earlier, making nMR observable only in sub-100 nm wide nanowires.
 Since the presence of magnetic impurities is not obvious, the model\cite{epl75} for the enhancement of critical current 
 under {\it B} as observed in MoGe and Nb nanowires \cite{prl97} do not hold here, where deliberate magnetic impurities were introduced to get nMR.
 \par
 Though nMR has also been 
 predicted in disordered superconducting wires \cite{andreevprl}, the contribution responsible for nMR within this 
 model is exponentially small, which is not the case in our experiment. It can also be thought of that the low-{\it B}
 nMR is due to the suppression of weak localization \cite{wl1,wl2} by magnetic field. But a closer look at our nMR results reveals some facts which are contradictory to the weak localization proposition \cite{wl3}. First, the change in {\it R} within the weak localization picture is only 2-3\%  \cite{wl1,wl2,wl3}, whereas we observed more than 20\% change of nMR over the {\it B} range. Second, the {\it T} variation of the phase coherence length, $L_{\phi}$. It is determined from the broadening ($\Delta B$) of nMR peak width around {\it B}=0. It is expected within weak localization model, that $\Delta B$ will increase with {\it T}, leading a decrease in $L_{\phi}$ \cite{wl3}. We observe, on the other hand, the peak gets sharper (see  figure \ref{figdRdT} or \ref{figdRdW}(d)) as {\it T} is increased. Because of the above stated features, weak localization correction to the conduction failed to explain our results, suggesting some other mechanism being operated here. 
 \par 
 Another explanation of this nMR at low {\it B} in sub 100 nm wide wires might be related to suppression of the non-equilibrium charge imbalance process at normal-superconducting (N-S) boundaries, as previously observed in Sn stripes\cite{jltp33} and narrow Al loops\cite{prb40}. In narrower samples, due to high current density phase-slip centers created and works as N-S boundaries.  At N-S interface the superconducting energy gap $\Delta(T)$ retrieve itself fully over the coherence length $\xi(T)$ of the superconductor. In presence of an external bias current for $\frac{\Delta}{k_{B}T}<1$, the excitations from N propagate into S result a quasi-particle current in the superconductor. The length scale of the quasi particle current is known as charge-imbalance relaxation length, $\lambda_{Q}$. This gives rise to finite boundary resistance corresponding to length  $\lambda_{Q}$, for each such boundary. In presence of {\it B},  $\lambda_{Q}$ decreases with increase in {\it B}, hence, the corresponding boundary resistance also decreased, lead to nMR in the system. The observation of nMR in sub 100 nm wide wires showed that suppression of non-equilibrium charge-imbalance processes by {\it B}, might play a key role in electronic transport below $T_{c}$, and on the contrary its effect on wider wires and the 2D film is negligible. Although the dimensions of tin micro bridges\cite{jltp33} is significantly larger than our wires, the Al loops\cite{prb40} are of the same dimensions. But these results were obtained and explained with charge imbalance theory at much higher {\it T} ($\geq$ 0.3 K) than ours (50 mK). The $T_{c}$ for Al is 1 K whereas in our case it is 1.8 K. Initially, the validity of charge imbalance mechanism is questionable for $T \ll T_{c}$, but later, of late, a low temperature limit for the charge imbalance process was proposed \cite{prb81} and experimentally observed\cite{prb83}.
 \par
 Although the phenomena of nMR is not totally understood yet, but it might be related to reduction or suppression of phase slip barrier ($\Delta_{0}$) or rate of activated phase slips as a aftermath of suppression of the charge imbalance length with {\it B}. 
 In the beginning, it appears contradictory, since the reduction of $\Delta_{0}$ will prompt the increase in {\it R}.
 It was argued in ref. \cite{pu41} that at low-{\it T}, reduction of $\Delta_{0}$ may lead to an increase in number of quasi-particles, 
 which in turn reduce {\it R}. Hence, the observed MR is related to the interplay between these two {\it B} dependent contributions. 
 It is quite feasible indeed, that over a certain region of parameter, the second mechanism wins over the first and {\it R} decreases with 
 increasing {\it B}. At high {\it B}, $\Delta_{0}$ is suppressed and {\it R} increased with {\it B} as expected. 
 However, it is very difficult to make any quantitative comparison with our experiments from this qualitative argument 
 and a strong theoretical modeling is still required.
 \section*{Summary and Conclusions}
 We perform magneto transport measurements at low-{\it T} environment in quasi-{\it 1D} structures of a:InO. 
 The nanowires were fabricated by {\it e}-beam lithography, followed by {\it e}-gun evaporation of high purity In$_{2}$O$_{3}$. 
 The {\it T}-dependence of {\it R} was explained in terms of available phase slip models. 
 Nanowires of width equal to or less than 100 nm, irrespective of length, show nMR near zero {\it B}. 
 This nMR was found to depend closely on both {\it T} and width of the nanowire. 
 This phenomena was qualitatively attributed due to an interplay between two field dependent contributions.
 \section*{Methods}
 \label{meth}
 Our data were obtained from studies of several nanowires of different widths ({\it w}) and lengths ({\it L}), but,
 we will mainly focus on the results for {\it L}=1 ${\mu }$m long wires of 3 different {\it w}. 
 These were initially 
 defined by {\it e}-beam lithography on a Si/SiO$_{2}$ wafer.  a:InO was deposited by {\it e}-gun evaporation of ultra-pure
 (99.999\%) In$_{2}$O$_{3}$ pellets in an O$_{2}$ atmosphere (P$_{O_{2}}$=1.05e-5 Torr). The thickness of deposition was 25 nm as measured {\it in-situ} by a quartz crystal monitor. Since a:InO has been extensively used 
 to investigate superconductor to insulator transition (SIT) in {\it 2D} \cite{murthyprl1,murthyprl2,maoznat,maozSR,adamprl}, we simultaneously prepared one reference film of size 50$\times$165 $\mu $m,  
 for comparative study. The electrically continuous a:InO films of this thickness are not granular in morphology.  
 The gold contacts were patterned via optical lithography, followed by metal evaporation and lift off.  
 The zero field measurements were performed in a He-3 cryostat and the magnetoresistance data were obtained by cooling the 
 devices in a dilution refrigerator. To ensure the reliability of measurements, the cryostat with well filtered coax lines were used. Experiment lines were twisted and fed through low-pass filters to eliminate the effect of external EM noise on measurements.  The transport measurements were conducted in a two-terminal configuration by low frequency (7.147 Hz) 
 ac lock-in techniques under current biased condition. The signal from the sample was amplified by a homemade differential preamplifier prior to be detected in lock-in amplifiers. The lock-in frequency was also chosen such that none of its higher order harmonics matches with the line frequency. All the instruments and the refrigerators were connected properly to a single ground to ensure that there exists no ground loop(s). While measuring in current biased condition, a standard requirement is to keep the bias current much smaller than the critical current to avoid hysteresis effects due to overheating. Additionally, in order to stay within the linear response regime, the measuring current should remain smaller than the characteristic scale $I_{0}=k_{B}T_{C}/\phi_{0}$ equal to few tens of nA for the majority of materials\cite{pr464}. We used 1 nA biasing current in all our measurements. The details of fabrication and measurements can be found in Ref.\cite{mitraprb}. Contact {\it R} was subtracted prior to any comparison.

 \section*{Acknowledgements}
 S.M. and G.C.T. thank Karen Michaeli, Sumilan Banerjee and Arjun Joshua 
 for fruitful discussions and department's submicron center for providing
 necessary support in device fabrication. 
 S.M. thanks VATAT program for post-doctoral fellowship support. 
 This work was supported by the Minerva Foundation with funding from Federal German Ministry 
 for Education and Research, and by a grant from the Israel Science Foundation.
 \section*{Author contributions statement}
 D.S. conceived the experiment,  S.M. and D.M. design the devices in e-beam lithography, S.M. fabricated the samples, S.M. and G.C.T. conducted the experiments in dilution refrigerator, S.M. G.C.T. and D.S. analyzed the results and wrote the paper.  All authors reviewed and commented on the manuscript.
 \section*{Additional information}
 The author(s) declare no competing financial interests.
\end{document}